\documentclass[paper]{ieice}
\usepackage[pdftex]{graphicx,xcolor}
\usepackage[fleqn]{amsmath}
\usepackage{newtxtext}
\usepackage[varg]{newtxmath}

\field{B}
\vol{105}
\no{12}

\title[Deep Learning for CSI]{Deep Learning-based Massive MIMO CSI Acquisition for 5G Evolution and 6G}
\authorlist{%
 \authorentry{Xin Wang}{n}{labelA}\MembershipNumber{}
 \authorentry{Xiaolin Hou}{n}{labelA}\MembershipNumber{}
 \authorentry{Lan Chen}{n}{labelA}\MembershipNumber{}
 \authorentry{Yoshihisa Kishiyama}{e}{labelB}\MembershipNumber{}
 \authorentry{Takahiro Asai}{e}{labelB}\MembershipNumber{}
}
\breakauthorline{3}
\affiliate[labelA]{The authors are with DOCOMO Beijing Communications Laboratories, Co. Ltd., No. 2
Kexueyuan South Road, Beijing 100191, China.} 
\affiliate[labelB]{The authors are with NTT DOCOMO, INC., 3-6 Hikari-no-oka, Yokosuka-shi, Kanagawa-ken, 239-8536, Japan.}

\received{2022}{1}{30}
\revised{2022}{5}{08}

\newcommand{\mat}[1]{\mathbf{#1}}
\newcommand{\revision}[1]{\textcolor{black}{#1}}

\begin{document}
\maketitle
\begin{summary}
   Channel state information (CSI) acquisition at the transmitter side is a major challenge in massive MIMO systems
  for enabling high-efficiency transmissions. 
   To address this issue, various CSI feedback schemes have been proposed,
   including limited feedback schemes with codebook-based vector quantization and explicit channel matrix feedback.
   Owing to the limitations of feedback channel capacity,
   a common issue in these schemes is the efficient representation of the CSI with a limited number of bits at the receiver side,
   and its accurate reconstruction based on the feedback bits from the receiver at the transmitter side.
   Recently, inspired by successful applications in many fields,
   deep learning (DL) technologies for CSI acquisition have received considerable research interest from both academia and industry.
   Considering the practical feedback mechanism of 5th generation (5G) New radio (NR) networks,
   we propose two implementation schemes for artificial intelligence for CSI (AI4CSI),
   the DL-based receiver and end-to-end design, respectively.
   The proposed AI4CSI schemes were evaluated in 5G NR networks in terms of
   spectrum efficiency (SE), feedback overhead, and computational complexity, and compared with legacy schemes.
   To demonstrate whether these schemes can be used in real-life scenarios, 
   both the modeled-based channel data and practically measured channels were used in our investigations.  
   When DL-based CSI acquisition is applied to the receiver only,
   which has little air interface impact, it provides approximately 25\% SE gain at a moderate feedback overhead level.
   It is feasible to deploy it in current 5G networks during 5G evolutions.
   For the end-to-end DL-based CSI enhancements, the evaluations also demonstrated their additional performance gain on SE,
   which is 6\% -- 26\% compared with DL-based receivers and 33\% -- 58\% compared with legacy CSI schemes.
   Considering its large impact on air-interface design, it will be a candidate technology for 6th generation (6G) networks,
   in which an air interface designed by artificial intelligence can be used. 
\end{summary}

\begin{keywords}
Channel state information, deep learning, and downlink MIMO transmission
\end{keywords}

\section{Introduction}\label{sec:intro}

Massive multiple-input multiple-output (MIMO)\cite{Marzetta2010} is a key technology for 5th generation (5G) New Radio (NR). Channel state information (CSI) on the transmitter side plays a key role in improving the spectrum efficiency of downlink transmissions in massive MIMO systems. Such information is necessary for downlink precoding schemes to improve the signal-to-noise ratio (SNR) at dedicated receivers and eliminate multiuser interference (MUI), which enables high-order multiuser MIMO (MU-MIMO) downlink transmissions.
\par

Therefore, diverse technologies for CSI acquisition have been studied in both academic research and the standardization of mobile networks. Limited feedback schemes with codebook-based vector quantization or explicit channel matrix/eigenvector feedback have been introduced based on legacy signal processing technologies. A common challenge of these approaches is to find an efficient way to represent CSI with a limited number of bits, which can be accommodated by the feedback channel with an acceptable overhead.
\par

Among them, limited feedback schemes\cite{limitedfb} have been widely used, especially for frequency division duplex (FDD) systems and time division duplex (TDD) systems without radio frequency (RF) calibration. For limited feedback schemes, the base station (BS) transmits reference signals to facilitate the user equipment (UE) estimating the downlink channel, and the UE quantizes the CSI with a specified codebook into a group of bits. These bits are fed back to the BS via the uplink feedback channel and the BS reconstructs the CSI based on them. The accuracy of the reconstructed CSI is constrained by the overhead introduced by the downlink reference signal and uplink feedback. For massive MIMO systems with a large number of antenna ports, more resources are necessary for transmitting downlink reference signals, and it is also difficult to quantize a large-scale CSI matrix into a limited number of bits without losing accuracy.
\par

For example, NR Release 15\cite{ts38214} specified two CSI limited feedback schemes, namely, Type I and Type II CSI. Type I CSI serves as a low-overhead scheme that quantizes the CSI into one of the codewords in the codebook, whereas Type II CSI uses a linear combination codebook, where multiple codewords can be selected from the codebook and the CSI is represented as a linear combination of selected codewords. Both the index of the selected codeword and quantized combination coefficients are sent back to the BS from UE. To control the feedback overhead, NR Type II CSI uses coarse quantization of combination coefficients and a large frequency domain granularity. Coarse quantization introduces severe nonlinear noise into feedback CSI. In addition, the large frequency domain granularity introduces information loss, which makes it difficult to reconstruct an accurate CSI with smaller granularity.
\par

As one of the key technologies in the new wave of artificial intelligence (AI), deep learning (DL) has been successfully used in many applications, such as image processing and natural language processing\cite{Simeone2018a}. Benefiting from the large-scale parallel computing power provided by modern processors, such as graphics processing units (GPUs) or tensor processing units (TPUs), DL networks trained with big data can usually find a better efficient solution to intractable problems compared with legacy optimization methods, especially for those in NP-hard families and those that are difficult to mathematically model \cite{Oshea2017,You2019a}. Deep neural networks can learn hidden mathematical models behind big data and fit nonlinear functions. One related example is image compression and super resolution\cite{Wang2019,Ahn2018,5466111,7780551}, where a deep neural network can learn the hidden structure inside images and then utilize this to compress and recover images or reconstruct high-resolution images based on their low-resolution versions.
\par

Similar to image compression or super-resolution, a complex model also hides behind the observed CSI, which is difficult to obtain practically and helpful for CSI acquisition. Motivated by this observation, DL technology has recently been introduced for CSI acquisition. In \cite{Wen2018b}, CsiNet was introduced as an end-to-end CSI compression and reconstruction method. Inspired by image super-resolution, SRCNN\cite{5466111} was used for channel estimation based on IEEE 802.11ad reference signals as a receiver enhancement scheme\cite{8761105}. Our previous work adopted an advanced super-resolution network VDSR\cite{7780551} to use cases of both end-to-end CSI acquisition\cite{chen2020deep} and receiver enhancements based on NR Type II feedback \cite{WangType2ai}.
\par

The progress of DL in CSI acquisition has attracted the attention of the industrial community. The 3rd Generation Partnership Project (3GPP) will initiate the study of this topic in the 2nd quarter of 2022\cite{rp213599}. The use cases of AI for CSI and their potential performance gain in practical mobile networks are the key objectives of this study. To address this issue, we introduce our AI for CSI (AI4CSI) use cases and evaluate their performance gains in NR networks. The contributions of this study are as follows.
\begin{itemize}
   \item Propose an enhanced end-to-end AI4CSI network based on the CsiNet in \cite{Wen2018b} with better performance, and extend our AI4CSI receiver in \cite{WangType2ai} to explicit feedback.
   \item Investigate the spectrum efficiency (SE), feedback overhead, and computational complexity of legacy and DL-based CSI feedback methods in a practical NR network, including Type I and II limited feedback schemes specified in current NR standards, explicit feedback schemes, and DL-based CSI for receiver enhancements (AI4CSI Rx) and end-to-end enhancements (AI4CSI E2E).
   \item Investigate the performance of AI4CSI schemes with measured channel data to verify their feasibility for practical deployments. 
\end{itemize}
Our study demonstrates the potential gain of AI4CSI schemes in a practical mobile system with limited feedback channel capacity. When AI4CSI is applied only to the receiver, which has little air-interface impact, it improves the accuracy of CSI reconstruction and requires fewer downlink reference signals. The performance gain in the SE of the downlink transmissions is approximately 25\%. Therefore, it is feasible for deployment in current mobile networks during network evolution. For AI4CSI E2E schemes, we show their significant performance gain, especially for moderate feedback overhead, and that it is feasible with practically measured channel data. This shows the prospects of applying an AI/DL defined air-interface for CSI in future networks, such as the 6th Generation (6G) mobile networks, where AI will be a native component of the networks.
\par 

The remainder of this paper is organized as follows. Section \ref{sec:model} introduces the proposed system.
Section \ref{sec:method} describes the legacy and the proposed AI4CSI schemes.
Performance evaluations with model-based channel data and practical channel measurements
are presented in Sections \ref{sec:results} and \ref{sec:realch}.
Finally, section \ref{sec:con} concludes the paper.
\par

Notations: Vectors (lower case) and matrices (upper case) are presented in boldface. $(\cdot)^\mathrm{T}$,
$(\cdot)^\mathrm{H}$ and $(\cdot)^\mathrm{-1}$ denote the transpose, conjugate transpose, and inverse, respectively.
\par

\section{5G NR Systems}\label{sec:model}

We studied the DL-based CSI feedback and reconstruction in a typical 5G NR mobile network. Following LTE specifications, NR also uses OFDM-based waveform and multiplexing schemes\cite{ts38200}, in which the time-frequency domain resources are divided into resource blocks (RBs), each of which comprises 14 OFDM symbols in the time domain and 12 subcarriers in the frequency domain. These time-and frequency-domain resources make up the resource grid of an OFDM system, where all channels and reference signals (RSs) are mapped onto the resource grid and then transmitted.
\par

\begin{figure}[h]
   \centering
   \includegraphics[width=2.3in]{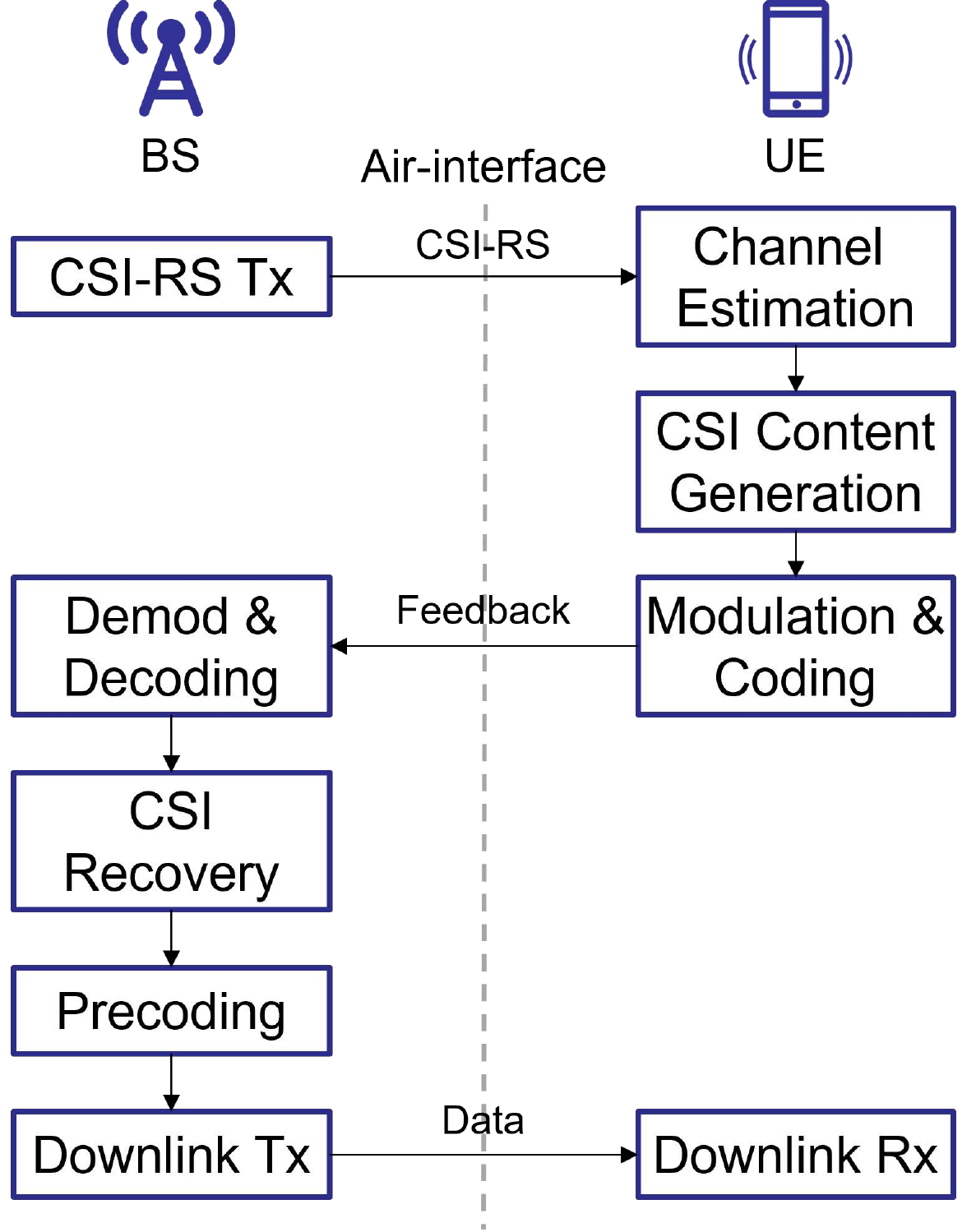}
   \caption{The diagram of CSI feedback and downlink transmissions.}
   \label{fig:overview}   
\end{figure}
\begin{figure}[h]
   \centering 
   \includegraphics[width=2.7in]{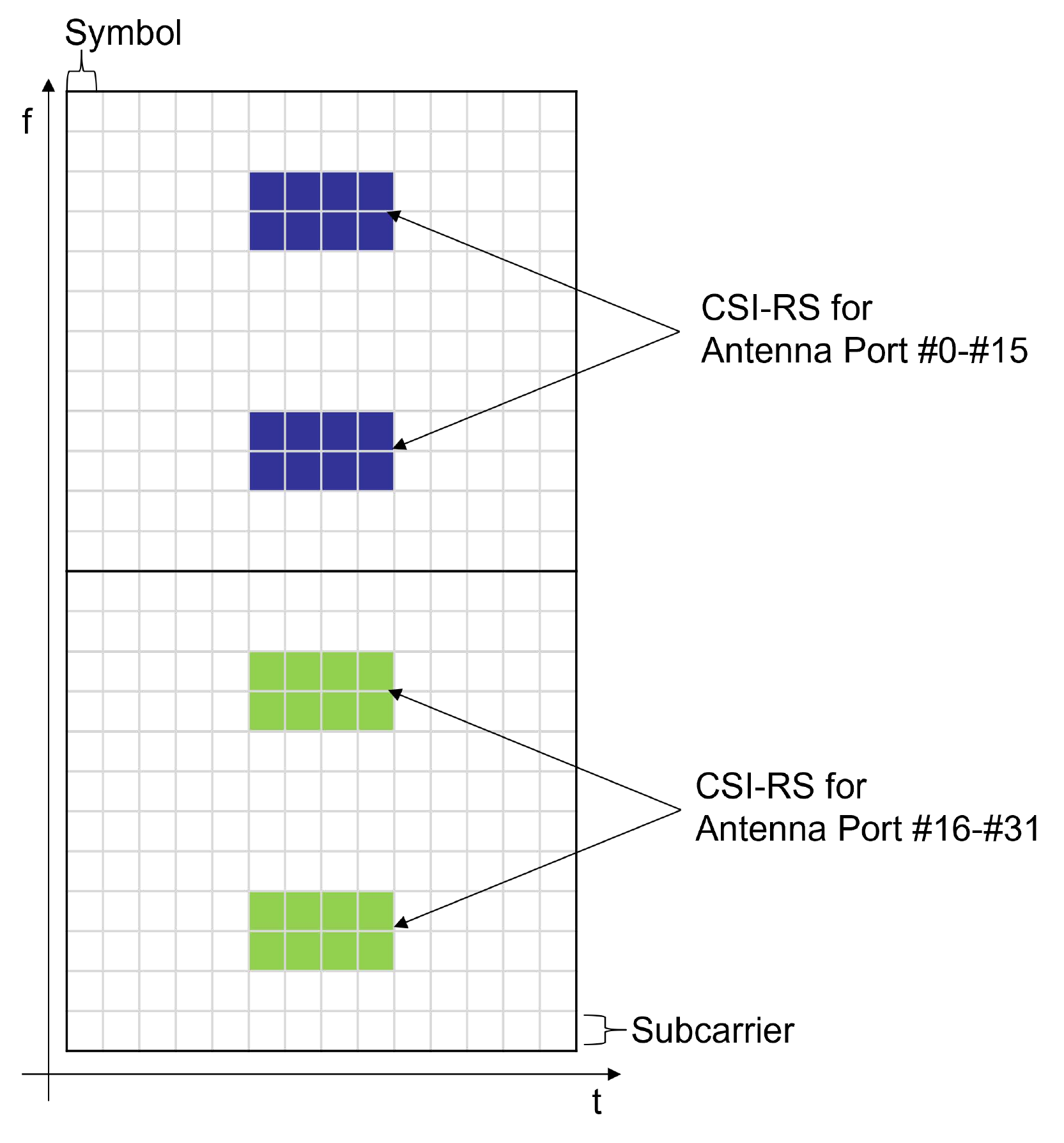}
   \caption{An example mapping pattern of NR CSI-RS.}
   \label{fig:csirs}   
\end{figure}

A typical closed-loop downlink transmission procedure with CSI feedback is shown in Fig. \ref{fig:overview}. In an NR system, CSI-RSs are transmitted from the BS to the UE to facilitate UE downlink channel estimation\cite{ts38211}. The current NR specifications support up to 32 antenna ports for CSI-RS transmissions. The signals of different ports are multiplexed using a combination of frequency-division multiplexing (FDM), time-division multiplexing (TDM), and code-division multiplexing (CDM). The frequency-domain density of NR CSI-RS is 1 or 0.5 resource elements per port per RB. An example of CSI-RS on an OFDM resource grid is shown in Fig. \ref{fig:csirs}.
\par
 
UEs estimate the downlink channel based on the received CSI-RS and then generate CSI feedback, which is transmitted to the BS via uplink feedback channels. When CSI is available at the BS, it can use typical downlink MIMO precoding, such as zero-forcing\cite{zfbd} to transmit data to multiple UEs with spatial multiplexing.
\par

Feeding downlink CSI back to BS efficiently has been discussed for decades , since MIMO technology was introduced into mobile systems. In the next section, we first introduce the basic signal model of the mobile system and then introduce several typical CSI acquisition schemes, legacy, and AI-based schemes.
\par
 
Suppose that there are $N_t$ antenna ports in the BS to serve $M$ UEs in a cell,
and there are $N_r$ antenna ports for each UE.
For an OFDM waveform with $K$ subcarriers, a general received CSI-RS at the UE can be written as
\begin{equation}
   \mat{r}_{m,k}=\mat{H}_{m,k}\mat{s}_{RS,k}+\mat{n}_{m,k},
\end{equation}
where $m=1,\cdots, M$ is the index of UE and $k$ is the index of the subcarriers where CSI-RS is transmitted.
$\mat{H}_{m,k}\in \mathcal{C}^{N_r\times N_t}$ denotes the downlink channel in the
$k$-th subcarrier, $\mat{s}_{RS,k}\in\mathcal{C}^{N_t\times N_t}$ is the transmitted CSI-RS signal,
and $\mat{n}_{m,k}\in\mathcal{C}^{N_r\times N_t}$ denotes the thermal noise of the receiver.
Without loss of generality, we assume that the CSI-RSs for different antenna ports are orthogonal.
Therefore, $\mat{s}_{RS,k}$ is a diagonal matrix in which the transmitted signals of CSI-RS appear as diagonal elements.
\par

After receiving CSI-RS, UEs estimate the channel based on the received signal and obtain the estimated channel 
$\hat{\mat{H}}_{m,k}, \forall k=1,\cdots,K$.
Although CSI-RS has a low frequency-domain density, UEs can use interpolation methods to obtain channels for all subcarriers,
for example, zero- or first-order interpolation as a simple implementation.
or MMSE interpolation using the known channel statistics.
\par

NR specifications define the frequency-domain feedback granularity as a subband
comprising several RBs.
For subband feedback, UEs can calculate the subband equivalent channel as
\begin{equation}
   \hat{\mat{H}}_{m}^{(s)}=\mat{v}_{N_r}\left(\sum_{k=k_s}^{k_e} \hat{\mat{H}}_{m,k}^{\mathrm{H}}\hat{\mat{H}}_{m,k}\right),
\end{equation}
where $s$ is the index of the subband, $k_s$ and $k_e$ are the start and end indices of subcarriers for a subband,
and $\mat{v}_{N}(\cdot)$ denotes $N$ eigenvectors corresponding to $N$ largest eigenvalues of a matrix.
\par

The CSI is usually decompensated into several components for separate feedback. Channel gain is usually reported as a factor of the channel quality indicator (CQI), which is the signal-to-interference-and-noise ratio (SINR) derived from the UE. The UE may also report a rank indicator (RI) that indicates the expected MIMO transmission layer number according to channel conditions.
\par

Among these components, the one representing the channel matrix is the most important. Because the channel gain is reported implicitly in CQI, the normalized channel matrix, which can be called channel direction information (CDI), should be reported to the BS from UE. There are many schemes that report CDI to BS explicitly or implicitly, which is the focus of our study and will be discussed in Section \ref{sec:method}.
\par

After receiving feedback from the UEs, the BS can reconstruct the channel matrix 
and prepare the downlink transmissions.
Define the channel matrix recovered by the BS as $\tilde{\mat{H}}_{m,k}$.
The downlink channel of all serving UEs is then
\begin{equation}
   \tilde{\mat{H}}_{k}=\left[
      \tilde{\mat{H}}_{1,k}^{\mathrm{H}}, \tilde{\mat{H}}_{2,k}^{\mathrm{H}}, \cdots, \tilde{\mat{H}}_{M,k}^{\mathrm{H}} \right]^\mathrm{H},
\end{equation}
where the BS can compute the downlink precoder $\mat{P}_{k}$ by following the steps in \cite{zfbd}.
For a simplified example, when the UE has one mounted antenna ($N_r=1$), the dimension of each $\tilde{\mat{H}}_{m,k}$ is $1\times N_t$.
The zero-forcing precoder can be obtained by
\begin{equation} \mat{P}_k=\tilde{\mat{H}}_{m,k}^{\mathrm{H}}\left(\tilde{\mat{H}}_{m,k}\tilde{\mat{H}}_{m,k}^{\mathrm{H}}\right)^{-1}.
\end{equation}
\par

The received signal of downlink transmissions for all UE antennas on $k$-th subcarriers is then
\begin{equation}
   \mat{r}_{k} = \mat{H}_{k}\mat{P}_k\mat{s}_{D,k}+\mat{n}_{k},
\end{equation}
where $\mat{H}_k=[\mat{H}_{1, k}^{\mathrm{H}},\cdots, \mat{H}_{M, k}^{\mathrm{H}},]^{\mathrm{H}}$,
$\mat{s}_{D,k}\in\mathcal{C}^{M \times 1}$ represents the data transmitted to the UE,
and $\mat{n}_{k}$ denotes the thermal noise of the receiver.
\par

\section{CSI Acquisition Schemes}\label{sec:method}

As discussed in Section \ref{sec:model}, the most important part of the CSI feedback is CDI, whose dimension increases with the number of antenna ports, and which becomes a large-scale matrix in massive MIMO systems.
\par

In legacy signal processing, the processing on CDI is categorized as a quantization or compression scheme. The objective is to represent the CDI through a limited number of bits without losing much information. Many schemes have been discussed over the decades, since MIMO technologies have been used in mobile systems. We introduce two types of legacy schemes in Section \ref{sec:method:legacy} and several DL-based schemes in Section \ref{sec:method:ai}.
\par

\subsection{Legacy CSI Acquisition Schemes}\label{sec:method:legacy}

\subsubsection{Limited Feedback of Precoding Matrix}

NR uses a limited feedback scheme that quantizes the precoding matrix expected by the UE into an index called the precoding matrix index (PMI) and reports it to the BS instead of reporting the CDI. The BS can then use the reported PMI directly to downlink single-user MIMO (SU-MIMO) transmissions. However, for MU-MIMO transmissions, the original CDI is required for the BS to calculate the MIMO precoder. Fortunately, the precoder matrix selected by the UE is usually highly correlated with the downlink channel, which means that we can treat the PMI as a representation of the CDI and reconstruct the channel matrix from it.
\par

The detailed procedure for NR feedback and recovery is defined in \cite{ts38214}. The UE first calculates the feedback content based on the subband equivalent channel $\hat{\mat{H}}_{m}^{(s)}$ and sends the feedback content to the BS. Here, we consider NR Type II feedback, which uses a two-stage linear-combination codebook. The BS can reconstruct the subband level feedback channel based on the UE CSI feedback, which can be represented as
\begin{equation}
   \tilde{\mat{H}}_{m}^{(s)}=\mat{W}_{m,1}\mat{W}_{m,2}^{(s)},
\end{equation}
where $\mat{W}_{m,1}$ consists of several DFT base vectors selected by the UE to represent the CSI, which is wideband feedback according to NR specifications, and $\mat{W}_{m,2}^{(s)}$ includes combination coefficients for subband $s$. If all base vectors are included in $\mat{W}_{m,1}$ and $\mat{W}_{m,2}^{(s)}$ is unquantized, there is no information loss in the PMI. However, to control the feedback overhead, only a subset of the base vectors is selected, and both the amplitudes and phases of the elements in $\mat{W}_{m,2}^{(s)}$ are quantized into a few bits, which introduces severe quantization noise.
\par

The reconstructed channel is then used for downlink transmissions. Because the feedback granularity is subband, the downlink precoder usually has to be calculated with the same granularity. Both quantization noise and large granularity degrade the SE of the downlink transmissions.
\par

\subsubsection{Explicit Feedback}

Another legacy CSI approach is explicit CDI feedback since itself is more expected for MU-MIMO precoding. CDI or its equivalent forms can be reported to the BS directly in explicit feedback schemes.
\par

As a typical explicit feedback scheme, we consider direct CDI feedback with scalar quantization in this study. The real and imaginary parts of the nonzero elements in $\hat{\mat{H}}_{m,k}$ are quantized into a specified number of bits. Considering the feedback overhead constraint in practical systems, there are typically insufficient resources to transmit all nonzero elements in $\hat{\mat{H}}_{m,k}$. Therefore, subband level feedback is also used to control the feedback overhead.
\par

Note that the vector quantization schemes used for implicit feedback can also be used for explicit feedback. In such a case, a similar performance is expected because quantization noise is the dominant factor in the performance.
\par

\subsection{Deep Learning based CSI Schemes}\label{sec:method:ai}

\subsubsection{DL-based CSI Reconstruction at Receiver (AI4CSI Rx)}

With the constraint of feedback overhead, limited information can be transmitted and received via feedback channels. To obtain a more accurate reconstructed CSI from limited information, the reconstruction algorithm should have detailed information on the inter-structure of the channel, which depends on the detailed radio propagation procedure. Practical over-the-air channels usually consist of many multi-path components, each of which is composed of many rays. If radio propagation can be mathematically described and accurately estimated, the CSI can be reconstructed accurately as well. However, it is a challenging task to model and estimate radio propagation using legacy signal processing technologies. The practical model is hidden behind the observable channels. By training with big data, DL networks can learn the hidden structure from a large number of channel samples and utilize the information it learns to reconstruct the CSI. Therefore, we consider using DL to reconstruct CSI from the inaccurate feedback of UEs.
\par

DL networks have been used for image super-resolution, which reconstruct a high-resolution image from its low-resolution copy\cite{Wang2019}. During the training stage of networks, the mapping between low-resolution and high-resolution images or the inner structure information can be learned using a large set of training data. A typical super-resolution network can process an image with multiple channels, improve the resolution of the images, and filter noise. For both implicit and explicit feedback, the feedback CSI is also two-dimensional data with multiple channels, such as real and imaginary parts or amplitude and phase parts. The CSI with large feedback granularity can be viewed as a low-resolution copy, and quantization noise is introduced, which should be filtered.
\par

\begin{figure}[h]
   \centering
   \includegraphics[width=3.0in]{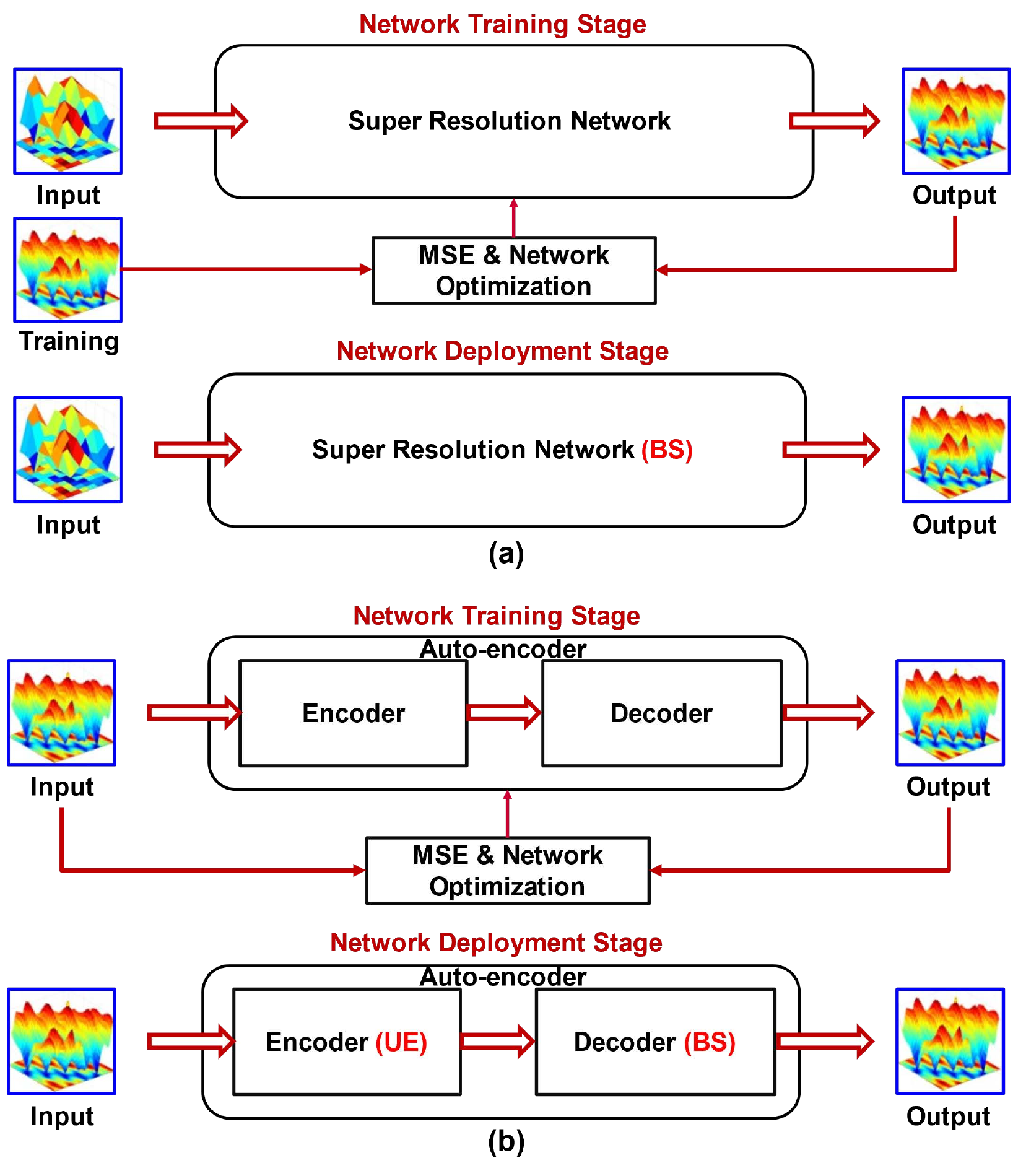}
   \caption{System structure for CSI reconstruction during training stage and deployment stage. (a) AI4CSI Rx. (b) AI4CSI E2E.}
   \label{fig:network}  
\end{figure}

Motivated by the similarity between CSI reconstruction and image super-resolution, we adopted image super-resolution networks to reconstruct the CSI with NR Type II feedback in \cite{WangType2ai}. General diagrams of the networks in the training and deployment stages are presented in Fig. \ref{fig:network} (a), respectively.

The proposed AI4CSI Rx network is deployed on the BS side. Because both NR and explicit feedback introduced in Section \ref{sec:method:legacy} can be viewed as low-resolution versions of CSI, the proposed network can accept either of them as inputs. After receiving UE feedback, the BS can reconstruct the channel matrix with any legacy interpolation method and then convert the complex numbers into real ones. The real and imaginary parts of the channel matrix are stored separately in the tensors and treated as two channels of the input data for the DL network.

Among the image super-resolution networks, we selected the VDSR network\cite{7780551} to reconstruct the CSI. It is a very deep convolutional neural network (CNN) with 16--20 convolution layers, using  which inner structure information is extracted from the training data. As the output of the network is an enhanced version of the input, VDSR also applies a global residual network architecture that is widely used for image processing.

During the training stage, we used a perfect channel matrix as the tag to train the entire network. The mean square error (MSE) between the output of the network and the perfect channel matrix was calculated and then backpropagated to update the coefficients of the CSI reconstruction network. During this stage, the CSI reconstruction network learns the structure between the low-resolution CSI and the expected perfect CSI. After training, the network was deployed to the BS.

\begin{figure}[h]
   \centering
   \includegraphics[width=3.0in]{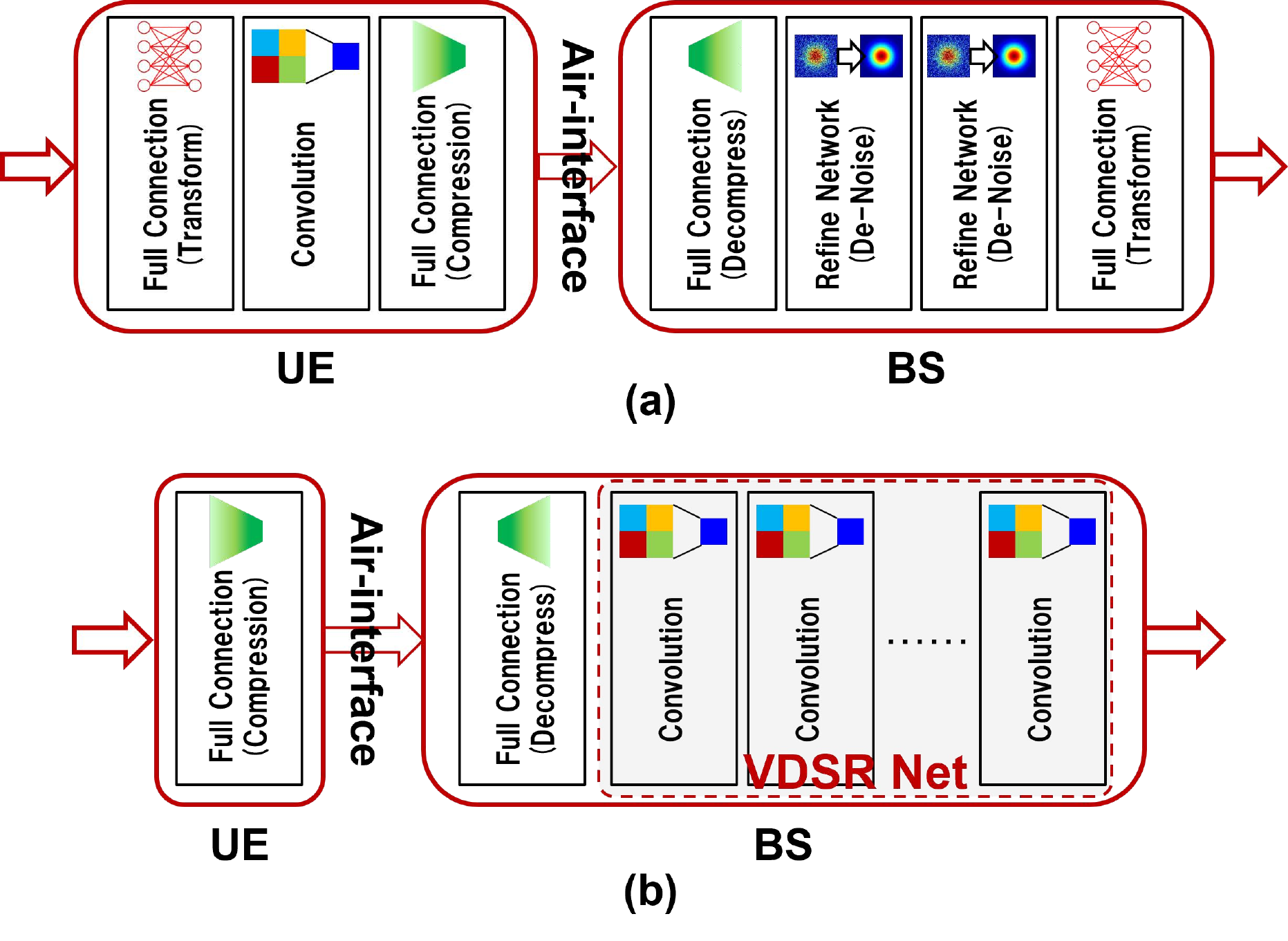}
   \caption{Network architectures of AI4CSI networks (a) The balanced design. (b) The Lean UE design.}
   \label{fig:ai4csi}  
\end{figure}

We consider two alternatives to AI4CSI Rx. One is AI4CSI Rx (Type II), the CSI reconstructed from NR Type II feedback, and the other is AI4CSI Rx (Explicit), the CSI reconstructed from the explicit feedback. The performance of both alternatives is investigated in Section \ref{sec:results}.
\par

\subsubsection{DL-based End-to-end CSI Acquisition (AI4CSI E2E)}

As introduced in Section \ref{sec:intro}, DL-based end-to-end enhancements have received considerable research interest in recent years. The ultimate objective of communication systems is to transmit information from one end (transmitter) to another (receiver) with minimal errors. For CSI feedback and reconstruction, the objective is to let the BS reconstruct a perfect CSI with feedback information from the UE. Existing feedback schemes are usually designed heuristically because it is difficult to jointly optimize all the factors in a typical feedback scheme, such as codebook selection, quantization method, quantization levels, and quantized bit allocation. DL provides an alternative approach in which DL networks design UE feedback and BS reconstruction, which is an end-to-end solution.
\par

In AI/DL technologies, an auto-encoder based on self-supervised learning has an end-to-end structure. The input of the network is also used as training data to train the network output. Therefore, the network is trained to generate outputs that are similar to the inputs. The autoencoder can be divided into two parts: an encoder and decoder. The last layer of the encoder, that is, the output layer of the encoder, can be viewed as a compression layer. Its output is the feedback content generated by the auto-encoder, the overhead of which can be controlled by restricting the number of nodes in this layer. Under such overhead constraints, the task of the decoder network is to recover the input with minimal error.
\par

Based on this feature of the auto-encoder, existing works use it for CSI feedback and reconstruction \cite{Wen2018b}. A diagram of the autoencoder-based CSI schemes is shown in Fig. \ref{fig:network} (b). On the UE side, the encoder network can be used to encode CDI into a small-scale vector, which is then quantized and transmitted to the BS. Subsequently, the BS can apply the decoder network to recover the CDI as shown in the deployment stage in Fig.\ref{fig:network} (b). The output of the encoder is a vector of real numbers that should be compressed and tends to be uniformly distributed. We then use scalar quantization to quantize each element of the vector into bits for feedback.
\par

Following the principle of the auto-encoder and pioneering work on its application to CSI feedback\cite{Wen2018b}, we consider two types of end-to-end CSI feedback and reconstruction networks. One is an enhanced network based on CsiNet \cite{Wen2018b}, the architecture of which is shown in Fig. \ref{fig:ai4csi} (a).
\par

Our proposed enhanced network considers a balanced design between the encoder and decoder, similar to CsiNet, in which the computational complexity of the encoder and decoder is comparable. Inspired by the fact that most data compression schemes transform the original data into a proper domain where the transformed data are sparse, we add a full connection layer before the original version of CsiNet to emulate the domain transformation operation. Unlike the domain transformation in existing compression schemes, the transformation in DL networks is obtained by training with big data rather than a predefined one. It is expected that the DL network can find a better domain to compress CSI than any domain used in legacy compression schemes. An inverse domain transformation layer is also added at the decoder side.
\par

Another AI4CSI E2E network to be evaluated has a lean-UE design. For AI4CSI E2E, the encoder is deployed in the UE and the decoder is deployed in the BS. Therefore, the complexity of the encoder is negligible to avoid increasing the UE processing complexity. Motivated by this, we consider a network with a lean UE design introduced in \cite{chen2020deep}. The architecture of the network is illustrated in Fig. \ref{fig:ai4csi} (b). For the lean UE network, only one necessary compression layer is maintained at the encoder side. On the decoder side, after the linear layer to recover the dimension of the CSI, super-resolution networks are still used to recover the full-resolution CSI and filter the noises.
\par

The two implementations introduced in this subsection are named AI4CSI E2E (balanced) and AI4CSI E2E (lean UE).
\par

\section{Performance Evaluations with Simulated Channels}\label{sec:results}

\subsection{Simulation Setup}
We evaluate the performance of all the proposed algorithms introduced in Section \ref{sec:method:legacy}
and Section \ref{sec:method:ai} in a typical NR network.
The major simulation assumptions are listed in Table \ref{tab:assumptions}, and follow those used in the study of NR systems.

In our simulations, an array with 32 antenna elements was mounted on the BS with a $4\times 4\times 2$ layout (horizontal $\times$ vertical $\times$ polarization), and one antenna was mounted on each UE. We consider four UEs that periodically report their CSI every 5 ms. After collecting their CSI reports, the BS transmits downlink data using ZF precoding\cite{zfbd} with the CSI it receives, which is obtained in different ways, as introduced in Section \ref{sec:method}.

In our evaluations, we considered different feedback configurations with different frequency-domain granularity and quantization accuracy. For NR CSI feedback schemes, the feedback overhead differs from these configurations, as shown in Table \ref{tab:feedback}. Note that we extended the range of the parameters specified in the current NR specification\cite{ts38214} to generate feedback content with different overheads.
\begin{table}[h]
   \centering
   \caption{Simulation Assumptions}
   \label{tab:assumptions}
   \begin{tabular}{|c|c|}
      \hline
      Parameter & Values\\
      \hline\hline
      Carrier Frequency & 4 GHz \\
      \hline
      Bandwidth & 100 MHz \\
      \hline
      \revision{Subcarrier Spacing} & \revision{30 kHz} \\
      \hline
      \revision{Subcarrier Number} & \revision{3276 (273 RBs)} \\
      \hline
      Channel Model & 3GPP NR CDL \\
      \hline
      BS Antenna Array & 32 $(4\times 4\times 2)$ \\
      \hline
      UE Antenna & 1 \\
      \hline
      UE Number & 4 \\
      \hline
      \revision{UE velocity} & \revision{3 kmph} \\
      \hline
      \revision{Maximum Doppler frequency} & \revision{11 Hz} \\
      \hline
      Feedback Period & 5 ms \\
      \hline
      SNR & 25 dB \\
      \hline
      Feedback Overhead & 10 kbps --- 8.8 Mbps (per UE)\\
      \hline
      CSI-RS Configuration & 1/2 --- 1/64 RE per port per RB\\
      \hline
   \end{tabular}
\end{table}
\begin{table}[h]
   \centering
   \caption{NR Type I/II Feedback Configurations}
   \label{tab:feedback}
   \begin{tabular}{|c|c|c|c|c|}
      \hline
      Overhead & Subband & Codeword & Phase & Amplitude \\
      (kbps) & Number & Number & Quantization & Quantization \\
      \hline\hline
      10 & 18 & 1 & 3 bit & 3 bit \\
      \hline
      94 & 18 & 4 & 3 bit & 3 bit \\
      \hline
      330 & 36 & 6 & 3 bit & 3 bit \\
      \hline
      570 & 54 & 6 & 4 bit & 3 bit\\
      \hline
      1100 & 108 & 6 & 4 bit & 3 bit\\
      \hline
   \end{tabular}
\end{table}

For explicit channel matrix feedback, the feedback overhead is controlled by feedback granularity, The density of the CSI-RS is also adjusted according to the feedback granularity. In the current NR specification, the frequency domain density of the CSI-RS can be as low as $1/2$ resource elements (RE) per port per RB. In our evaluations, in addition to $1/2$ RE per port per RB, we further reduced the density of the CSI-RS and lowered the density to $1/128$ RE per port per RB. The evaluation configurations for explicit channel feedback are shown in Table \ref{tab:exp}. \revision{For all cases shown in Table \ref{tab:exp}, we use 5 bits to quantize both the real and imaginary parts of the complex numbers.} We also evaluated the performance of implicit feedback methods with different CSI-RS densities to test their super-resolution capability.

The feedback overhead of AI4CSI E2E scheme can be controlled by configuring the encoder network.
\revision{In this work, we consider that each output of the compression layer is quantized into 5 bits, the same as the bit width used for explicit feedback schemes.}
Therefore, we can flexibly adjust the overhead of AI4CSI E2E schemes according to requirements.

We used the NR CDL channel models specified in \cite{tr38901} to generate the training, verification, and test datasets for all schemes.
\revision{There are multiple models with different power-delay profiles defined in \cite{tr38901}. To avoid overfitting of the DL network, our training dataset consisted of channel samples of different models and root mean square (RMS) delay spreads (DS). More specifically, we generated the datasets by mixing CDL-A, CDL-B, and CDL-C models, each of which consists of 23--24 multipath components with different powers and delays. The RMS DS of CDL models can be scaled to simulate different practical channel conditions. In the simulations, the training data set has 100'000 channel samples, each of which is generated by randomly selecting one of the three CDL models, and then randomly selecting its RMS DS from short (30 ns), nominal (100 ns), and long (300 ns). The DL network can learn more propagation conditions from these training datasets. We also generate verification and test data sets, each of which has 1'000 samples, following the same way. }

\begin{table}[h]
   \centering
   \caption{Explicit Channel Feedback Configurations}
   \label{tab:exp}
   \begin{tabular}{|c|c|c|}
      \hline
      Overhead & CSI-RS Density & Port Number \\
      (kbps) & (RE/RB/Port) & \\
      \hline\hline
      17 & 1/128 & 8 \\
      \hline
      68 & 1/128 & 32 \\
      \hline
      270 & 1/64 & 32 \\
      \hline
      550 & 1/32 & 32 \\
      \hline
      1100 & 1/16 & 32 \\
      \hline
   \end{tabular}
\end{table}
\begin{figure}[h]
   \centering
   \includegraphics[width=3.0in]{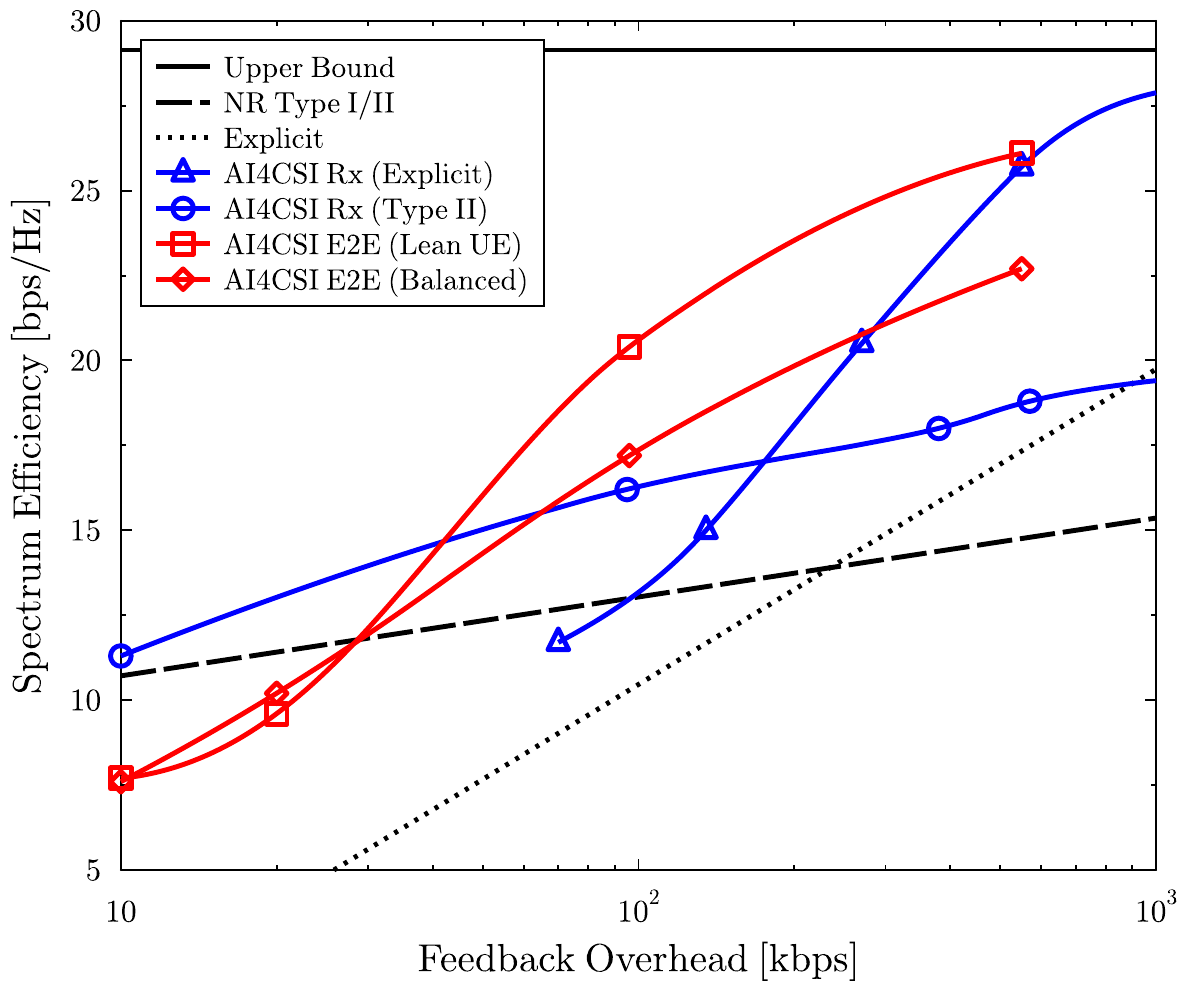}
   \caption{SE of schemes with different feedback overhead.}
   \label{fig:simu1}  
\end{figure}
\begin{figure}[h]
   \centering
   \includegraphics[width=3.0in]{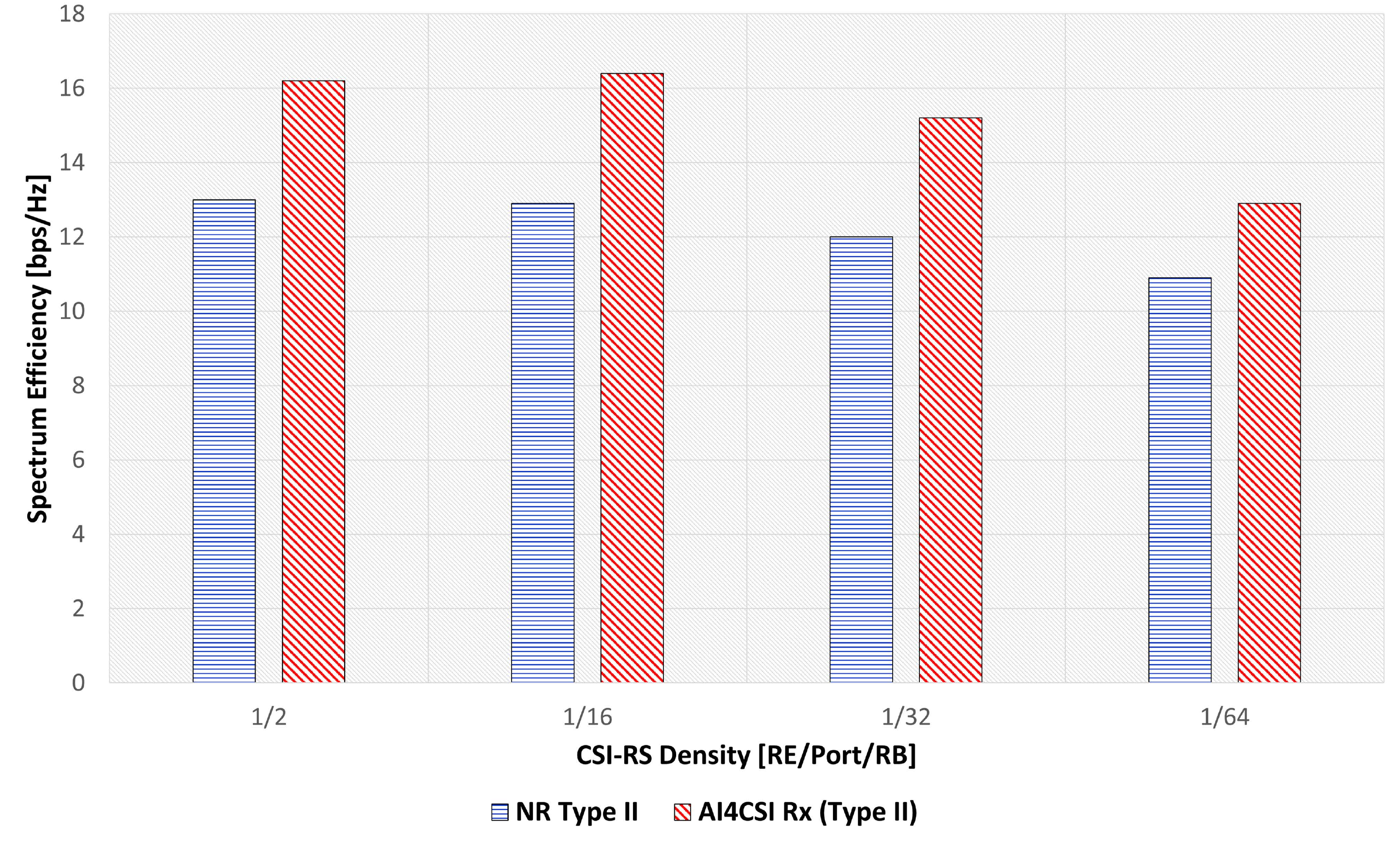}
   \caption{SE of implicit feedback schemes with different CSI-RS density.}
   \label{fig:simu2}  
\end{figure}

\subsection{Spectrum Efficiency}

Fig. \ref{fig:simu1} shows the performance of the baseline and AI4CSI schemes. First, we discuss the tendency of legacy signal processing algorithms. As shown in the figure, the achievable SE of the two legacy schemes tends to be linear with the feedback overhead. For NR Type II feedback, the vector quantization used in the schemes exhibits better performance at a low overhead level. However, explicit feedback can provide a more accurate CSI with medium to high overheads. There is a large area on the upper left of both curves, where a better trade-off between the SE and feedback overhead can be achieved. This is the objective of AI4CSI schemes.

It is illustrated in Fig. \ref{fig:simu1} that AI4CSI Rx schemes can recover the channel matrix with higher accuracy compared with legacy schemes. Using the reconstructed CSI, there is an SE gain of approximately 25\% when AI4CSI Rx is used with NR Type II feedback. 

In Fig. \ref{fig:simu2}, we plotted the SE of schemes with different CSI-RS densities. It is shown that AI4CSI Rx provides $18\%$---$27\%$ SE gain. When the density of CSI-RS is $1/64$ per port per RB, which is $1/32$ of the current specified density, AI4CSI Rx provides similar spectrum efficiency as the legacy schemes with current CSI-RS density. This demonstrates its capability of working with very low frequency-domain granularity, which is benefitted from the capability of super-resolution networks.

From the results shown in Fig. \ref{fig:simu1}, AI4CSI Rx applied on explicit feedback can also improve the SE compared with the baseline schemes. When explicit feedback can provide a sufficiently accurate channel matrix, the super-resolution network can recover CSI accurately enough for MU-MIMO transmissions, which makes the SE approach the ideal CSI upper bound. Similar to the performance of legacy implicit and explicit feedback schemes, the DL-based approach with explicit channel matrix feedback is not superior to the same network with implicit feedback. The two curves cross in the moderate feedback overhead region. This hints at considering detailed feedback schemes and CSI-RS configurations according to the acceptable feedback overhead that can be accommodated in uplink channels.

Both AI4CSI E2E networks can improve the SE-feedback overhead tradeoff when the feedback overhead is moderate, that is, approximately 100 kbps per UE. The performance gain on the SE is 33\% -- 58\% compared with the NR Type II feedback. However, these two networks cannot outperform the NR feedback at very low feedback overhead levels, that is, when the feedback overhead is approximately 10--30 kbps for each UE. Under such feedback overhead conditions, the number of nodes in the compression layer of the network is constrained to 10--30. The network must find a scheme to represent a large-scale channel matrix with only 10 real numbers. This is difficult and requires more channel data to train the network for this objective.

Between the two different end-to-end networks, the network with the lean UE design exhibits better performance in most cases. Although the complexity of the UE network is significantly reduced, we used a large and very deep CNN on the BS side. With the price of complexity, it achieves better performance than a balanced design. This complexity issue is discussed in the following subsections.

\subsection{Computational Complexity}

Following the simulation assumptions in Table \ref{tab:assumptions}, we analyzed the computational complexity of each scheme. We used the number of float operations per second (FLOPS) to denote the computational complexity of the algorithms.
\revision{The complexity is evaluated when the feedback overhead is 550 kbps (570 kbps for NR Type II feedback because its feedback overhead cannot be adjusted flexibly). The computational complexity of operations for pre-and post-processing, such as basic matrix calculations and fast Fourier transforms, was also evaluated.}
\begin{table}
   \centering
   \caption{The computational complexity the schemes under investigation.}
   \label{tab:complexity}
   \begin{tabular}{|c|c|c|}
      \hline  
      Scheme & BS Complx. & UE Complx. \\
       & (GFLOPS) & (GFLOPS) \\
      \hline\hline
      NR Type II & $<0.1$ & 0.9 \\
      \hline 
      Explicit & $<0.1$ & $<0.1$ \\
      \hline 
      AI4CSI Rx (Type II) & 243 & 0.9 \\
      \hline 
      AI4CSI Rx (Explicit) & 243 & $<0.1$ \\
      \hline
      AI4CSI E2E (Lean UE) & 243 & 0.5 \\
      \hline
      AI4CSI E2E (Balanced) & 3 & 1 \\
      \hline
   \end{tabular}
\end{table}

\revision{The computational complexities for each scheme are listed in Table \ref{tab:complexity}. Because the UE needs to calculate and select the codeword from the codebook with NR CSI feedback schemes, where operations such as vector correlation or even singular value decomposition may be involved, the complexity of NR Type II feedback on the UE side cannot be ignored. However, the complexity on the BS side is trivial because it only needs to combine the vectors reported by UE. However, for practical systems, we usually expect that computational complexity can be on the BS side with a lean UE design.}

By enlarging the network scale as well as the complexity on the BS side, AI4CSI E2E schemes can have a lean UE design and still have a performance gain compared with legacy methods. The UE can easily compress the CSI with a network that has only one layer. Even when considering a balanced complexity design, the UE complexity is comparable with existing schemes, and the BS complexity can be dramatically reduced.

\section{Performance Evaluations with Real-life Channel Measurements}\label{sec:realch}

\revision{We evaluated the performance of AI4CSI schemes with channels generated by models used in 3GPP, which is a kind of stochastic spatial correlation channel model. However, such channels are more deterministic than those in real life and can be easily managed by deep learning. The generalization performance of deep learning for communications, that is, whether it is still superior to legacy methods in practical deployments, is a major concern in using them in practical systems. In this section, we provide initial trials of AI4CSI schemes with real-life channel samples. Since AI4CSI E2E schemes have shown their potential gain with moderate feedback overhead in the simulations, we focus on their performance in the evaluations with real-life channels.}
\par

\begin{table}[h]
   \centering
   \caption{MSE of end-to-end CSI feedback and recovery with realistic channel measurements.}
   \label{tab:real}
   \begin{tabular}{|c|c|c|c|}
      \hline
      Scheme & BS Complx. & UE Complx. & SE \\
       & (GFLOPS) & (GFLOPS) & (bps/Hz) \\
      \hline\hline
      NR Type II & $<0.1$ & 0.9 & 5.7 \\
      \hline
      AI4CSI E2E (Balanced) & 2 & 1 & 14.5 \\
      \hline
      AI4CSI E2E (Lean UE) & 60 & 0.5 & 15.4 \\
      \hline
      AI4CSI E2E (Lean UE) & 240 & 0.5 & 20.9 \\
      \hline
      AI4CSI E2E (Lean UE) & 960 & 0.5 & 26.3 \\
      \hline
   \end{tabular}
\end{table}

Although there is no common dataset for AI-based physical layer design in mobile systems, we attempt to utilize some open datasets to test whether the AI4CSI schemes can be deployed in real-life scenarios. 
\revision{The channel measurements used for the evaluations shown in this section are introduced in \cite{csidataset}. The channels are measured in an indoor office scenario on the 3.5 GHz band with 100 MHz bandwidth. The MIMO configuration is $4\times 4$. The measured channel dataset contains 320'000 samples that are measured at different locations in the scenario. During the measurements, the receiver moved at a speed of 1.5m/s.}
\par

\revision{Considering that the characteristics of the real-life channel are different from the ones obtained with models, the AI4CSI networks were retrained with real-life channels. In practice, it is usually possible to collect channel samples from cells before the deployment of base stations and use them to train the network. We randomly extracted 1'000 channel samples from all 320'000 samples to create a verification dataset, and another randomly selected 1'000 samples to set up the test dataset. The remaining data were used as the training dataset. Similar to the simulations, we evaluated the SE of the AI4CSI E2E schemes to demonstrate their overall performance for downlink transmissions. }
\par

\revision{The network architecture of AI4CSI networks is the same as the ones used in the previous section. Because real-life channels are obtained with different MIMO configurations, the hyperparameters of the networks were adjusted to accommodate the dimensions of the data. Following the feedback configurations in \cite{csidataset}, we considered a 25.6 kbps feedback overhead (128 bits for each feedback) for all cases. The configurations of NR Type II CSI were also modified to generate similar feedback overheads. The subband number was configured to 10, the number of combined codewords was three, and the combination coefficients were quantized with 3 bits for amplitudes and 4 bits for phases.}
\par

\revision{During the trials, we noticed that it requires larger network models to handle the real-life channel measurements because the channel characteristics are more diverse than that generated by models. The propagation conditions, that is, line-of-sight or non-line-of-sight, and the distribution of multipath components on different positions in the scenario, are variant. Therefore, in addition to the networks used in Section \ref{sec:results}, we also consider more complex BS side networks with larger widths, that is, larger channel numbers of the convolution layers, to check if more complex networks can handle real-life data. Because we do not want to introduce additional complexity for the UE, the networks on the UE sides are unchanged. The computation complexity of the networks is shown in Table \ref{tab:real}, which was evaluated with the assumption of a 5 ms feedback period (200 feedbacks per second). For AI4CSI E2E (balanced) network, because it is not flexible enough to adjust the complexity owing to its complex network architecture, we only tested one network size configuration. For AI4CSI E2E (lean UE), the network on the BS side is a very deep convolution neural network whose width can be adjusted flexibly. We tested more cases of different complexities. The width of the convolution layers in the network was 64 when the complexity was 60 GFLOPS and 128 and 256 when the complexity was 240 GFLOPS and 960 GFLOPS, respectively. Note that the complexity of networks with the same width is smaller than that of the networks used in the previous section owing to the different configurations of MIMO and feedback overhead. }
\par

It is shown in Table \ref{tab:real} that AI4CSI E2E schemes are superior to NR Type II feedback schemes in terms of the SE. During the evaluations, we found that if more complex networks on the BS side are allowed, \revision{ that is, the computational complexity is increased to 240 or 960 GFLOPS from 60 GFLOPS }, the SE is continuously and rapidly improved by the complexity of the networks. This verifies the capability of AI4CSI schemes to handle practical data in real-life deployments, although the cost of computational complexity should be considered.
For the balanced design of the network, although there is no easy way to scale it, it still provides a better tradeoff between complexity and performance compared to the lean UE design. Because the information is lost after air interface transmission between the encoder and decoder, it is natural that more processing at the UE can improve to the final performance.
\par

In summary, from this trial with real-life channel data, we find that the AI4CSI E2E schemes can still work as they do with channels generated by the models. The feasibility of these schemes was verified.
\par

\section{Conclusion}\label{sec:con}
In this study, we investigated DL-based CSI acquisition schemes for massive MIMO with the practical settings of current 5G mobile networks. We investigated the SE, feedback overhead, and computational complexity of our proposed two types of DL-based schemes, that is, DL-based receiver enhancements and end-to-end design, as well as NR CSI feedback and explicit channel matrix feedback.
\par

The evaluations demonstrated the promising gain of DL-based CSI acquisition for 5G evolution and 6G. At a moderate feedback overhead level, that is, approximately 100 kbps per UE, the DL-based receiver enhancement introduces approximately 25\% performance gain on the SE of downlink transmissions, and the end-to-end design provides more performance gain on SE, up to 58\%, compared with legacy schemes. The performance gain of DL-based CSI acquisition schemes was also tested with a dataset consisting of channel measurements in a practical scenario, which verifies their feasibility for real-life deployments. As DL-based receiver enhancement has little impact on the air interface, it can be used in 5G networks with 5G evolution without much standardization effort. DL-based end-to-end CSI acquisition redefines the feedback air interface of the networks, where the feedback content is designed by the DL network itself. To obtain an additional performance gain, a new CSI feedback framework with native AI support, which enables CSI feedback and reconstruction designed by end-to-end DL networks, can be considered in future 6G networks.
\par

\bibliographystyle{ieicetr}
\bibliography{IEEEfull,iphy}

\end{document}